\documentclass{article}

\usepackage{graphicx}
\usepackage{psfig}
\usepackage{epsfig}
\usepackage[round]{natbib}

\title{Singular vector ensemble forecasting systems and the prediction of flow dependent uncertainty}

\author{Stephen Jewson\footnote{\emph{Correspondence address}: RMS, 10 Eastcheap,
London, EC3M 1AJ, UK. Email: \texttt{x@stephenjewson.com}}\\
RMS, London, United Kingdom\\
Maarten Ambaum and Christine Ziehmann}

\begin{document}

\newcommand{\bx}[1]{\fbox{\begin{minipage}{15.8cm}#1\end{minipage}}}

\maketitle

\begin{abstract}
The ECMWF ensemble weather forecasts are generated by perturbing the initial conditions
of the forecast using a subset of the singular vectors of the linearised 
propagator. Previous results show that when creating probabilistic forecasts 
from this ensemble better forecasts are obtained if the mean of the spread and
the variability of the spread are calibrated separately. 
We show results from a simple linear model that suggest that this may be a generic
property for all singular vector based ensemble forecasting systems based 
on only a subset of the full set of singular vectors.
\end{abstract}

\section{Introduction}

We are interested in the question of how to make forecasts of the distribution of 
future temperatures over time-scales of one or two weeks. 
The best predictors for this distribution come from numerical weather forecasting models, and, in particular,
from ensemble integrations of such models. 
The ensembles are generated by running the forecast model many times
from different initial conditions, and, in some cases, by using stochastic parameterisations. 
The precise methods used to generate the ensemble of
different initial conditions differ from one forecast system to another. For instance, ECMWF uses a method
based on perturbing the initial state using the singular vectors of the linearised propagator over a finite time
period~\citep{molteniet96} while NCEP uses the breeding vector method~\citep{tothk93}.

The predictors one can derive from these models are a mixture of information and error,
and it is non-trivial to convert these predictors into optimal probabilistic forecasts. 
The most straightforward
method used to derive a prediction of the future distribution of temperatures is to build a linear
regression model between the ensemble mean (as input) and the temperatures being predicted (as output).
Such a model gives a mean-square error minimising prediction of the temperature, as well as a prediction
of the uncertainty around that temperature. We will refer
to the regression model as a \emph{first generation} calibration model. It has been in use
since the 1970s (see, for example, \citet{leith}).

Second generation calibration models use more information from the ensemble than
just the ensemble mean in an attempt to predict flow dependent variations in the uncertainty
of the temperature prediction. 
Second generation models include the rank histogram~\citep{talagrandet97}, 
the best members method~\citep{roulstons03} 
and the spread-scaling method~\citep{jewson03g}. 
These models are similar in that
they calibrate the mean level of the uncertainty and the variability
of the uncertainty in the same way. This, it turns out, is not ideal, and as a result the second generation
models do not, generally, perform as well as linear regression. To understand why not one can
consider the case
in which the variability of the ensemble spread contains no information at all. In such a situation an effective
calibration method would ignore this variability and produce a forecast with a constant level of uncertainty
derived entirely from past forecast error statistics.
However, all of the second generation 
calibration models fail this test and would actually \emph{inflate} the variability in the uncertainty 
still further, because of the need to
inflate the mean level of spread.

The third generation of calibration models addresses this issue, and calibrates the mean and the variability of
the uncertainty in separate ways.  ~\citet{jewsonbz03a} describe such a model and 
show an example where the mean level of uncertainty needs to be 
increased while
the variability of the uncertainty needs to be decreased. 
Why should this be necessary? One explanation is that site
specific temperatures are affected by small-scale processes that increase the mean level of uncertainty,
but do not change the variability in the uncertainty. 
However, in this paper we will investigate another possibility: that it is the methods used to generate
the ensembles themselves, and in particular the truncated singular vector approach used at ECMWF,  
that result in ensemble forecasts that require that the mean and the variability of the 
uncertainty be calibrated separately. We approach
this question using a simple linear stochastic model of the prediction of forecast uncertainty.
This model allows us to calculate both the exact uncertainty and the uncertainty predicted by singular
vector methods, and to compare the two.

In section~\ref{ecmwf} we give a brief description of how the initial conditions are generated in the ECMWF
ensemble forecasting system. In section~\ref{model} we describe the simple model we will use to study
the properties of singular vector forecasting systems in general. In section~\ref{results} we present
our results and in section~\ref{discuss} we summarise.

\section{The ECMWF ensemble prediction system}
\label{ecmwf}

The method used at ECMWF to create an ensemble forecast can be described as follows.

\subsection{Step 1: linearisation of the propagator}

We write the entire atmospheric model as:
\begin{equation}\label{fx}
  \frac{dx}{dt}=F(x)
\end{equation}
where $x(t)$ is the atmospheric state and $F$ is a non-linear function representing the dynamics of the
model. We will assume henceforth that the model is perfect, and hence that $F(x)$ is also an accurate
representation of the non-linear dynamics of the atmosphere. We will only consider forecast errors that
arise due to errors in the initial conditions.

If we now consider a forecast made from the current state $x$, and write an initial condition error
as a small perturbation $e$ around $x$ we have:
\begin{eqnarray}\label{fx+e}
  \frac{d(x+e)}{dt}&=&F(x+e) \\\nonumber
                   &=&F(x)+\frac{dF}{dx}e+... 
\end{eqnarray}

Subtracting equation~\ref{fx} from equation~\ref{fx+e}, and ignoring higher order terms, gives:
\begin{eqnarray}
  \frac{de}{dt}&=&\frac{dF}{dx}e \\\nonumber
               &=& A(x(t))e
\end{eqnarray}

This is a linear equation for the development of initial condition errors in the forecast.

This can be solved to give:
\begin{equation}
  e(t)=\mbox{exp} \left( \int Adt \right) e_0
\end{equation}

Writing $B(t)=\int Adt$ gives:
\begin{equation}
  e(t)=\mbox{exp}(B(t)) e_0
\end{equation}

Expanding this gives:
\begin{equation}
  e= (1+B+...) e_0
\end{equation}

and ignoring higher order terms:

\begin{eqnarray}\label{linear}
  e(t)= (1+B(t)) e_0
\end{eqnarray}

This is now a linear equation which gives the forecast error at time $t$ in terms of the initial condition 
error at time 0.

\subsection{Step 2: creation of initial conditions}

The first 25 singular vectors of matrix $1+B$ are calculated.

\subsection{Step 3: creation of the ensemble forecast}

Positive and negative versions of each singular vector are propagated 
forwards using equation~\ref{fx} to give 50 forecasts.
The spread of these forecasts gives an indication of the uncertainty in 
the forecast.

\section{The model}
\label{model}

Our model for the process by which forecast errors in the ECMWF model develop is just equation~\ref{linear}: 

\begin{equation}
  e=(1+B)e_0
\end{equation}

where $e_0$ is a vector of initial condition errors, $e$ is a vector of final forecast errors, and $1+B$ is
a matrix representing the process by which the forecast errors grow.
For simplicity, we will study a 2 dimensional system.

We will assume that the matrix $B$ varies in time, to represent variations in 
the state of the atmosphere.
Our model for $B$ will be that each of the four matrix elements are independent
of the other elements, are independent of themselves in time, 
and are given by a standard normal distribution.

Our model for the real initial condition errors $e_0$ will be that these errors are
drawn from a bivariate normal distribution with correlation of zero.

Within this model the variations in the statistics of the distribution of forecast errors 
are driven by variations in the extent to which
the error propagator $1+B$ causes the initial condition distribution to grow or not.

\subsection{Generating the real forecast uncertainty}
\label{generating}

For each point in time (i.e.\ for each randomly generated $1+B$) 
we define the real forecast uncertainty by performing an ensemble of integrations
over 1000 values for $e_0$ sampled from the distribution of possible values
(the bivariate normal).
We take the resulting ensemble of 1000 values of $e$ and calculate the
distribution for the first element to represent observing a single variable.
The standard deviation of this distribution gives a
measure of the real uncertainty in the forecast (by definition).

\subsection{Predicting the forecast uncertainty using a full singular vector system}

We now imagine that we want to predict the uncertainty defined in section~\ref{generating} 
using the singular vector method.
We assume that we know the exact propagator for the forecast errors i.e.\ we also use
$1+B$ to predict the uncertainty. In real forecast systems, the process by which
forecast errors grow is not entirely understood. However, in our system we understand it
completely. 
 
We generate the ensemble prediction by calculating the two right singular vectors of $1+B$, 
and using them to initialise $e_0$ four times,
for positive and negative versions of each singular vector (mimicking the ECMWF initialisation system).
This generates an ensemble of
4 values for $e$ (which are the left singular vectors scaled by their singular values)
and the standard deviation of this ensemble gives the predicted uncertainty.

We then compare the temporal variations in the predicted uncertainty with the temporal
variations in the actual uncertainty. 

\subsection{Predicting the forecast uncertainty using the first singular vector}

We now imagine that we want to predict the uncertainty using the \emph{truncated} singular vector method.
This is closer to the system used at ECMWF, which is based on the first 25 singular vectors of a
system with many thousands of degrees of freedom. 
 
We now generate the ensemble prediction using only the \emph{first} singular vector of $1+B$.
We initialise, as before, using positive and negative versions of this singular vector.
This generates an ensemble of
2 values for $e$, and the standard deviation of this ensemble gives our prediction of the
uncertainty.

\subsection{Predicting the forecast uncertainty using the second singular vector}

Our third and final method for using singular vectors to predict the uncertainty uses only
the \emph{second} singular vector. Otherwise this method is identical to the second method.

\section{Results}
\label{results}

\begin{figure}[!htb]
  \begin{center}
    \scalebox{0.9}{\includegraphics{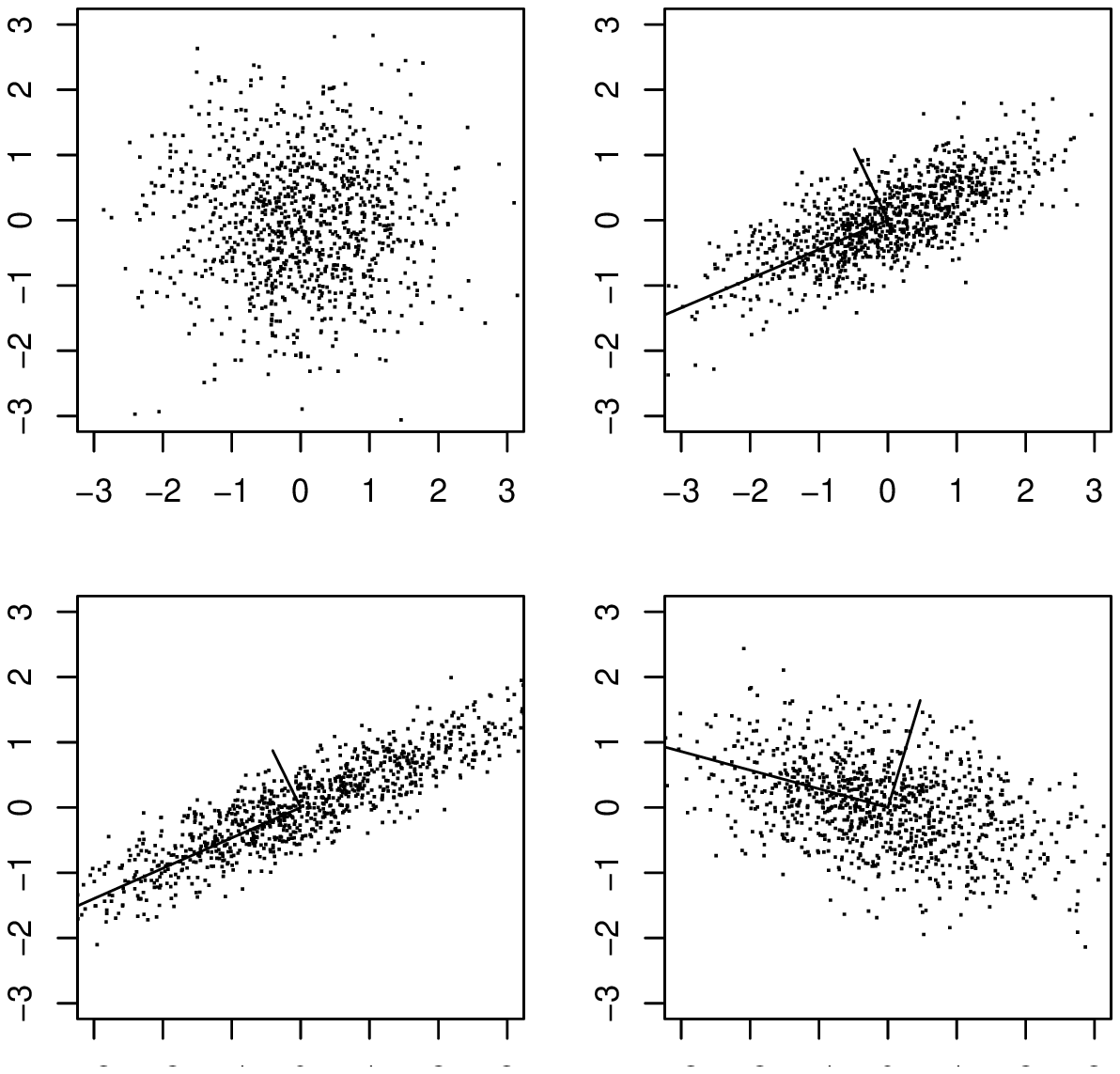}}
  \end{center}
 \caption{
Initial conditions, forecasts and left singular vectors scaled by their singular values...examples 1,2,3
} 
 \label{fig1}
\end{figure}

\begin{figure}[!htb]
  \begin{center}
    \scalebox{0.9}{\includegraphics{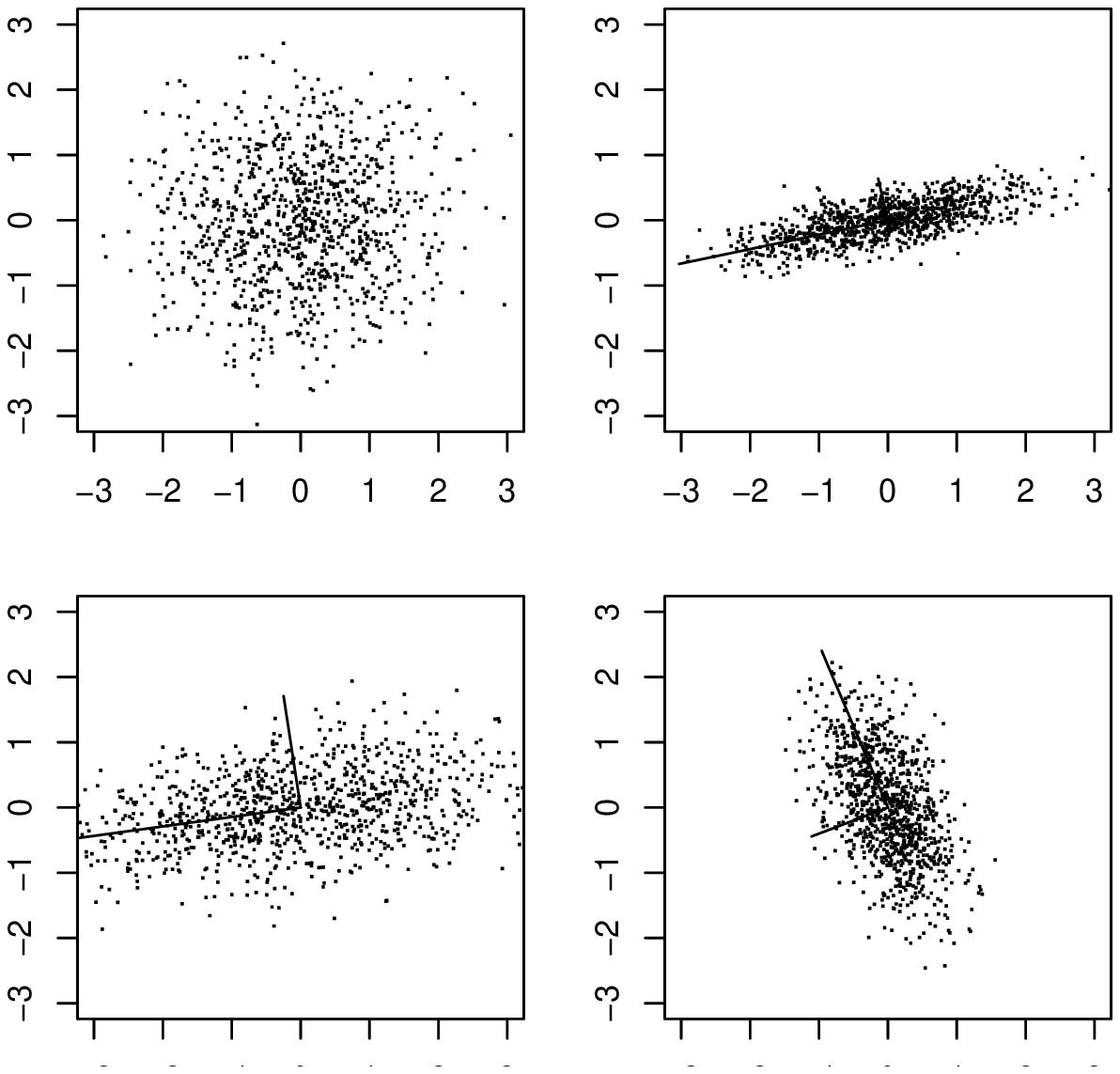}}
  \end{center}
 \caption{
Initial conditions, forecasts and left singular vectors scaled by their singular values...examples 4,5,6
} 
 \label{fig2}
\end{figure}

Figure~\ref{fig1} and figure~\ref{fig2} show examples of the initial conditions and
forecast errors from the random initial condition model, along with the left
singular vectors of the matrix $1+B$ scaled by their singular values, for six arbitrarily chosen forecast days.
 We see that the initial condition error ball
becomes an ellipse of final forecast errors, and that the singular vectors are
aligned with the principal axes of this ellipse, exactly as we would expect.

\subsection{Temporal variation of the real forecast uncertainty}

We now consider the temporal variations in the real uncertainty, generated from the ensemble of
initial conditions sampled from the bivariate normal distribution.

\begin{figure}[!htb]
  \begin{center}
    \scalebox{0.9}{\includegraphics{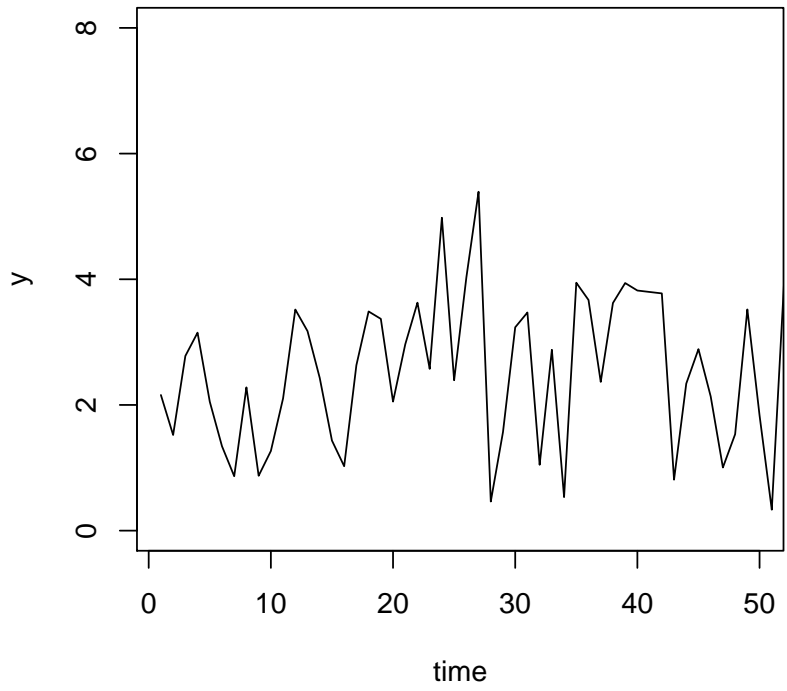}}
  \end{center}
 \caption{
Time series of the real uncertainty, generated using an ensemble of 1000 members based
on initial conditions from a bivariate normal distribution.
} 
 \label{fig3}
\end{figure}

Figure~\ref{fig3} shows a time series of 50 days of forecast uncertainty generated by the model.
We see that the forecast uncertainty varies in time. This is because of the random variations in the matrix
$B$, which mimic the flow dependent changes in the processes that control the growth of forecast errors. 

The top left panels of figure~\ref{fig8}, figure~\ref{fig9} and figure~\ref{fig10} show
the distribution of the real uncertainty. This distribution is repeated as a dotted line
in the other panels of these figures for purposes of comparison. 

\subsection{Temporal variation of the forecast uncertainty from the full singular vector model}

We now attempt to predict the temporal variations in the uncertainty using the full singular vector ensemble.
A scatter plot of the real uncertainty (horizontal axis) and the predicted uncertainty (vertical axis)
is shown in the lower left panel of figure~\ref{fig5}. We see a strong relation between the two, although
the predicted uncertainty is larger than the real uncertainty.
The lower right panel of figure~\ref{fig8} shows the distribution of predicted uncertainty (solid line), 
which is clearly too wide relative to the real uncertainty (dotted line).
We calculate the empirical correlation between this predicted uncertainty and the real
uncertainty: it is very close to 1.0. 
This shows the value of using singular vectors: we can avoid having to sample
the whole initial condition error ball by using vectors which efficiently span the forecast error space. 

We now consider calibration of the forecast, since the mean and the standard deviation of the uncertainty
prediction are both wrong (as shown by the distribution of the uncertainty in figure~\ref{fig8}).
We try a very simple calibration consisting of a scaling of the uncertainty forecast
(the "spread-scaling" method of~\citet{jewson03g}). Since we are
dealing with a linear system this is equivalent to scaling the initial condition singular vectors.
The effect of this calibration on the distribution is shown in the lower right panel of figure~\ref{fig9}.
We see that this simple calibration method succeeds in setting both the 
mean and the standard deviation of the uncertainty 
to be correct, even though there is only a single calibration parameter. A scatter plot of the 
calibrated forecasts is shown in figure~\ref{fig6}, lower left panel.

In summary, our full singular vector uncertainty forecast has a correlation with the real
uncertainty of one, and,
post-calibration, the mean and the standard deviation are correct. It is producing a perfect
forecast of the actual uncertainty. 

\subsection{Temporal variation of the forecast uncertainty predicted from the first singular vector}

We now consider the temporal variations in the uncertainty predicted using just the first singular vector.
This system is an analogy to the ECMWF prediction system, which uses a truncated set of singular vectors
to create an ensemble. 

\begin{figure}[!htb]
  \begin{center}
    \scalebox{0.9}{\includegraphics{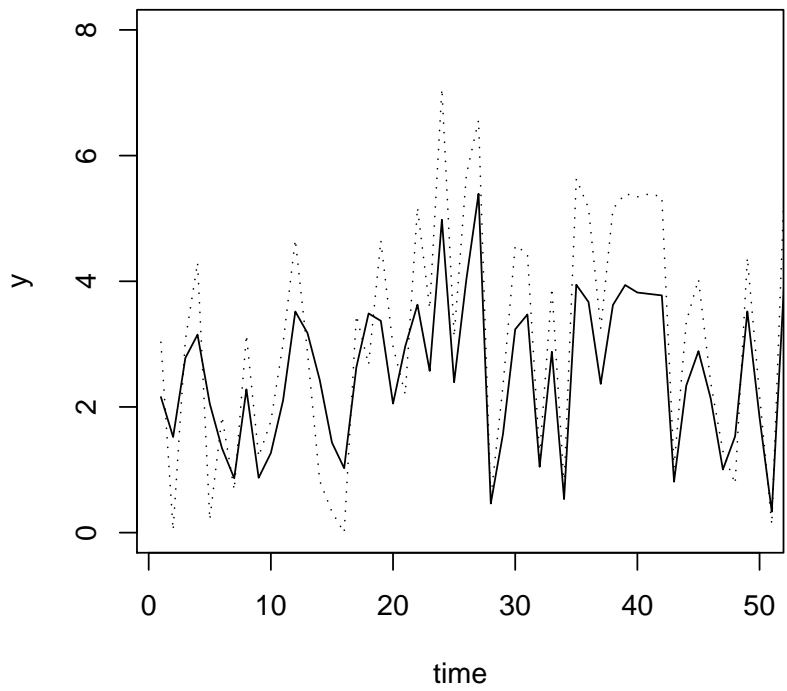}}
  \end{center}
 \caption{
Time series of the real uncertainty (solid line) 
with the uncalibrated prediction of the uncertainty from the first singular vector (dotted line).
} 
 \label{fig4}
\end{figure}

\begin{figure}[!htb]
  \begin{center}
    \scalebox{0.9}{\includegraphics{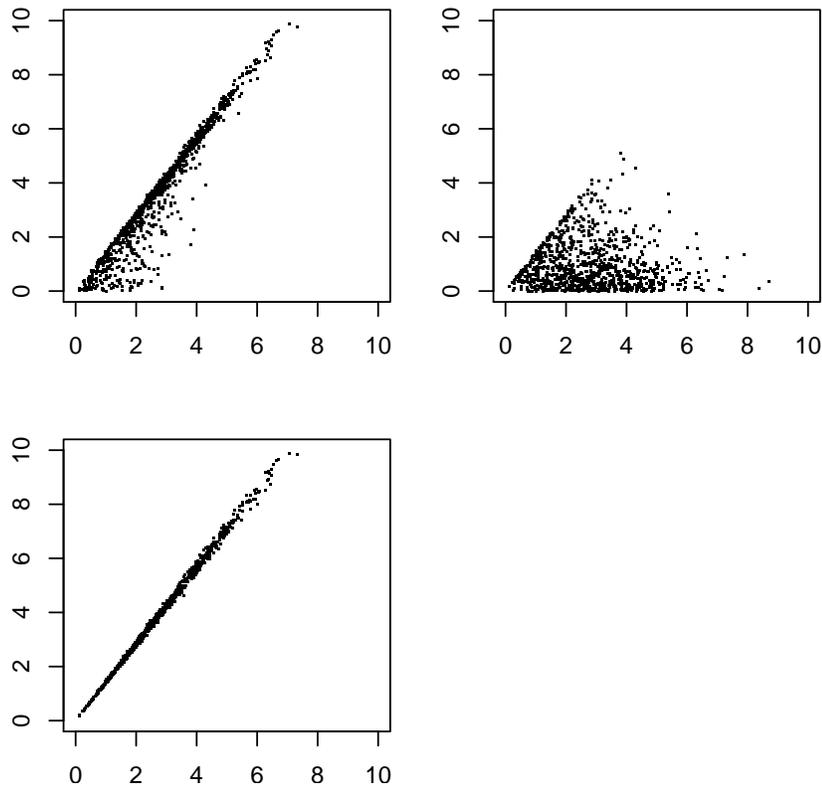}}
  \end{center}
 \caption{
Scatter between the real uncertainty (horizontal axes) and the 3 predictions of the uncertainty (vertical axes).
The top left panel shows the uncertainty predicted from the first singular vector, the top right panel shows
the uncertainty predicted from the second singular vector and the lower left panel shows the uncertainty
predicted from both singular vectors. All predicted uncertainties are uncalibrated.
In the top left panel we see that the predicted uncertainty is strongly related to the actual uncertainty.
All the points on the diagonal line are situations where the real uncertainty is dominated by the first singular
vector, and hence where the prediction using the first singular vector is a good one. The points below the
diagonal line correspond to situations where the first singular vector is less important (presumably because
it is more or less orthogonal to the observation axis) and where the second singular vector becomes important.
In these cases the predictions of uncertainty are poor because the second singular vector is not being used. 
In the top right panel we see much lower correlation. There is only a very weak diagonal line, corresponding
to these situations when the second singular vector dominates the uncertainty. 
The points below this line correspond to situations
where the first singular vector dominates. The real uncertainty in these situations is often large, but poorly predicted.
In the lower left panel we see a very high correlation between real and predicted uncertainty. 
There is a very small spread because
of sampling errors.
The slope of the diagonal line is not one: the predictions have a larger standard deviation than the real uncertainty.
}
\label{fig5}
\end{figure}

\begin{figure}[!htb]
  \begin{center}
    \scalebox{0.9}{\includegraphics{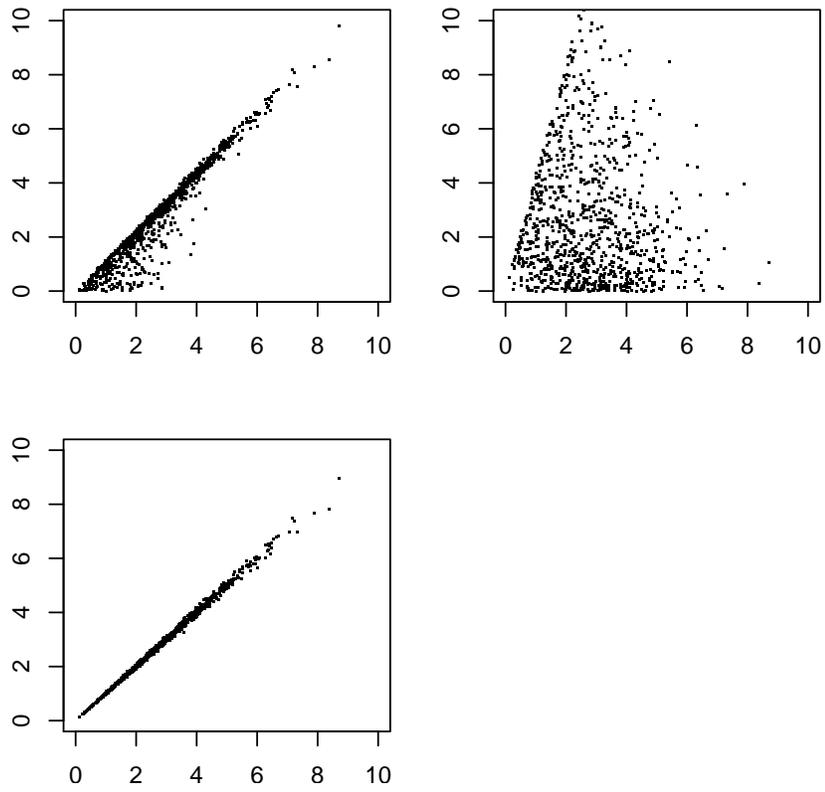}}
  \end{center}
 \caption{
As figure~\ref{fig5} but for the uncertainties calibrated using spread-scaling.  
In the lower left panel the slope of the line is now 1: only a single scaling is needed to calibrate the
forecast. 
}
\label{fig6}
\end{figure}

A time series of the predicted uncertainty is shown in figure~\ref{fig4} (dotted line) along with the
real uncertainty (solid line). The upper left panel of figure~\ref{fig5} shows a scatter plot of the
real and the predicted uncertainty. The upper right panel of figure~\ref{fig8} shows the distribution
of the predicted uncertainty.
The correlation between this predicted uncertainty and the actual uncertainty is approximately 0.95:
3 estimates of this correlation from a set of independent experiments are shown in table~\ref{table1}. 
From the correlation and the scatter plot we see that the use
of only a single singular vector is not as accurate as using the full set of singular vectors, as expected.

The mean and the variability of the predicted uncertainty are wrong, and, as before, we can 
attempt to calibrate them. 
This time, however, when we calibrate with a simple scaling designed
to correct the mean uncertainty, this does \emph{not} correct the standard deviation of the uncertainty
correctly, as is shown in the top right panel of figure~\ref{fig9}. 
A scatter plot of the results of this calibration are shown in the upper left panel of figure~\ref{fig6}.
The ratio of the standard deviation of the predicted uncertainty to the standard deviation of the 
actual uncertainty after the scaling calibration
is given in the second
column of table~\ref{table1}. We see that the calibrated forecast overestimates the variability of the
uncertainty. In a real forecast system this would lead to overprediction of extreme events. 

\begin{figure}[!htb]
  \begin{center}
    \scalebox{0.9}{\includegraphics{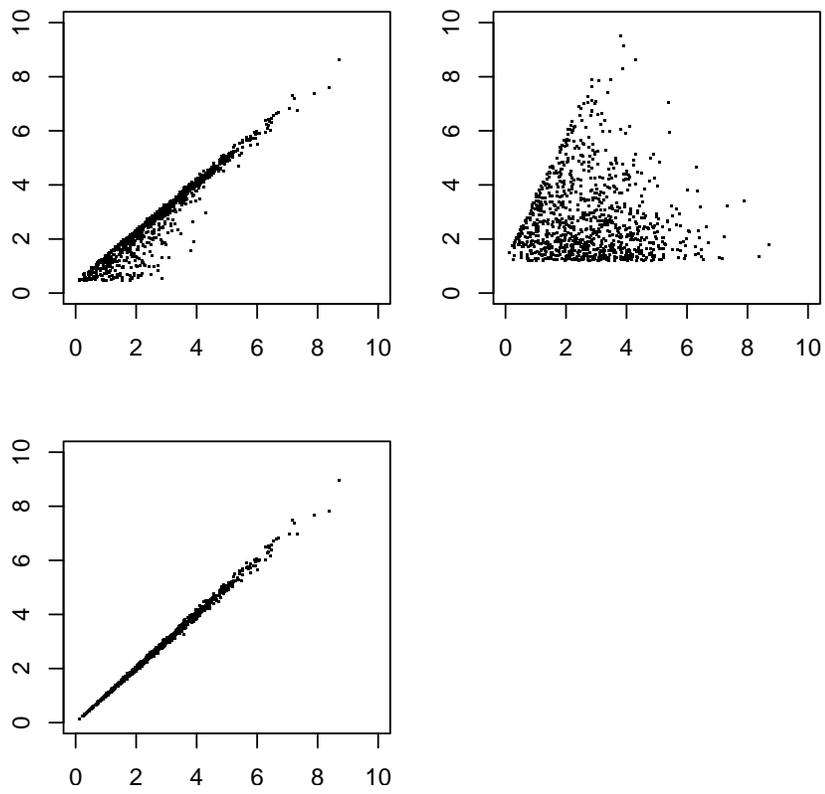}}
  \end{center}
 \caption{
As figure~\ref{fig5} and figure~\ref{fig6} but for the uncertainties calibrated using
a shift and a scaling.}
\label{fig7}
\end{figure}

\begin{figure}[!htb]
  \begin{center}
    \scalebox{0.9}{\includegraphics{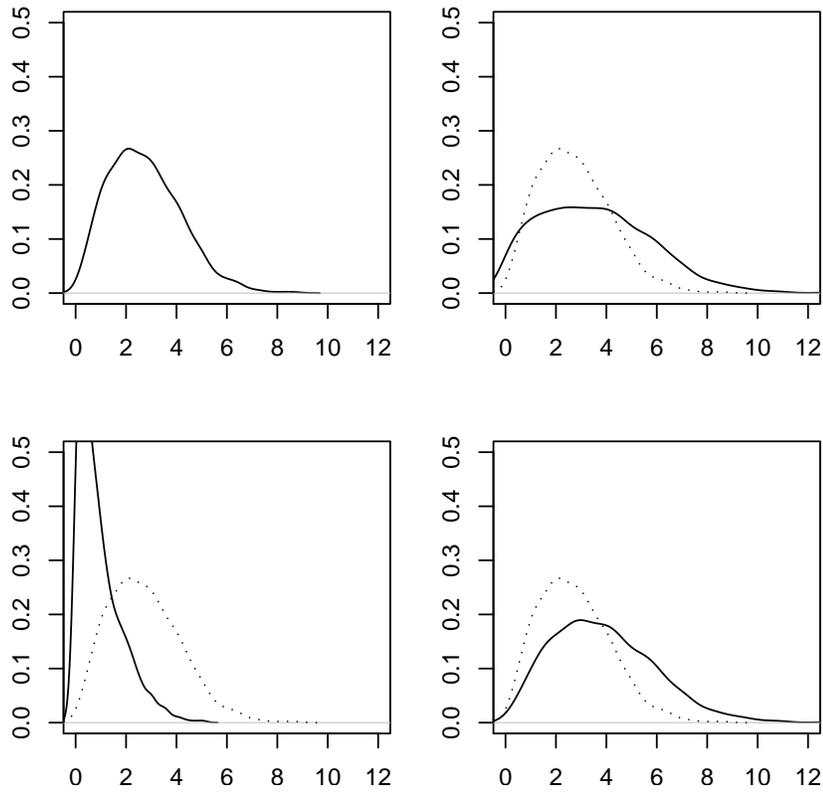}}
  \end{center}
 \caption{
Distributions of uncertainty, derived from 1000 samples using kernel smoothing.
The top left panel shows the distribution of the real uncertainty, and this curve is repeated in the
other panels as a dotted line.
The top right panel shows the distribution of uncertainty predicted using the first singular vector.
The lower left panel shows the distribution of uncertainty predicted using the second singular vector.
The lower right panel shows the distribution of uncertainty predicted using both singular vectors.
All three predicted uncertainties are uncalibrated.
We note that some of these curves show non-zero density for negative values. This is an artefact of
the kernel smoothing. 
In the top right hand corner we note that the predicted density (solid line) is much wider than the actual
density (dotted line), as is the case in the lower right hand panel.
}
\label{fig8}
\end{figure}

\begin{figure}[!htb]
  \begin{center}
    \scalebox{0.9}{\includegraphics{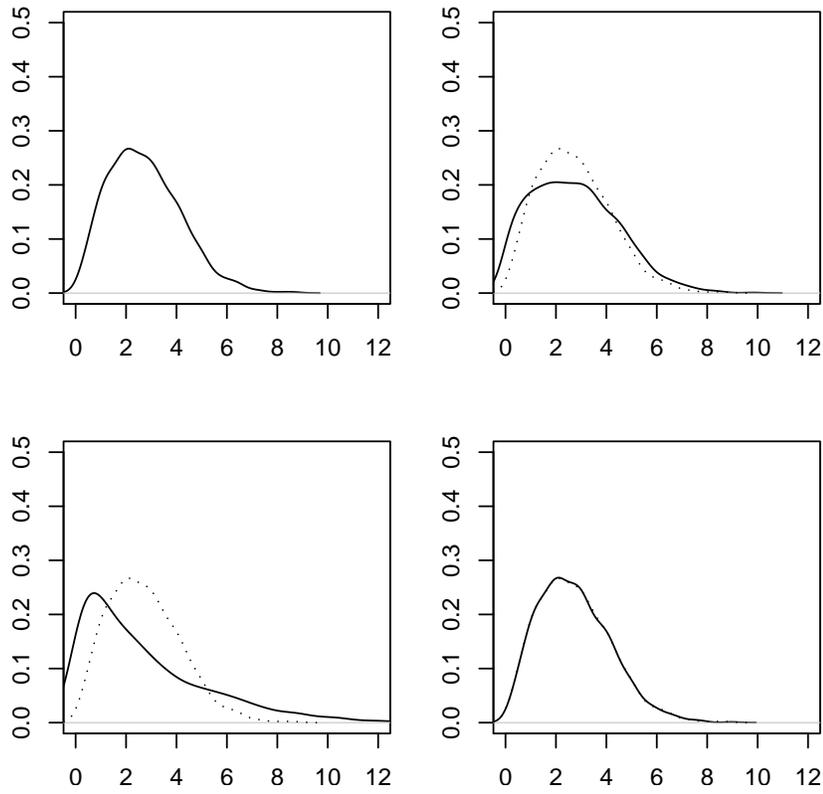}}
  \end{center}
 \caption{
As for figure~\ref{fig8} but for predicted uncertainties calibrated using spread-scaling.
In the lower right hand panel the calibration has rendered the predicted distribution more or less correct,
while in the top right hand panel the distribution is still too wide. This, we believe, is analogous to the 
overestimation of the variability of the uncertainty seen in real ensemble forecasts by~\citet{jewsonbz03a}.
}
\label{fig9}
\end{figure}

\begin{figure}[!htb]
  \begin{center}
    \scalebox{0.9}{\includegraphics{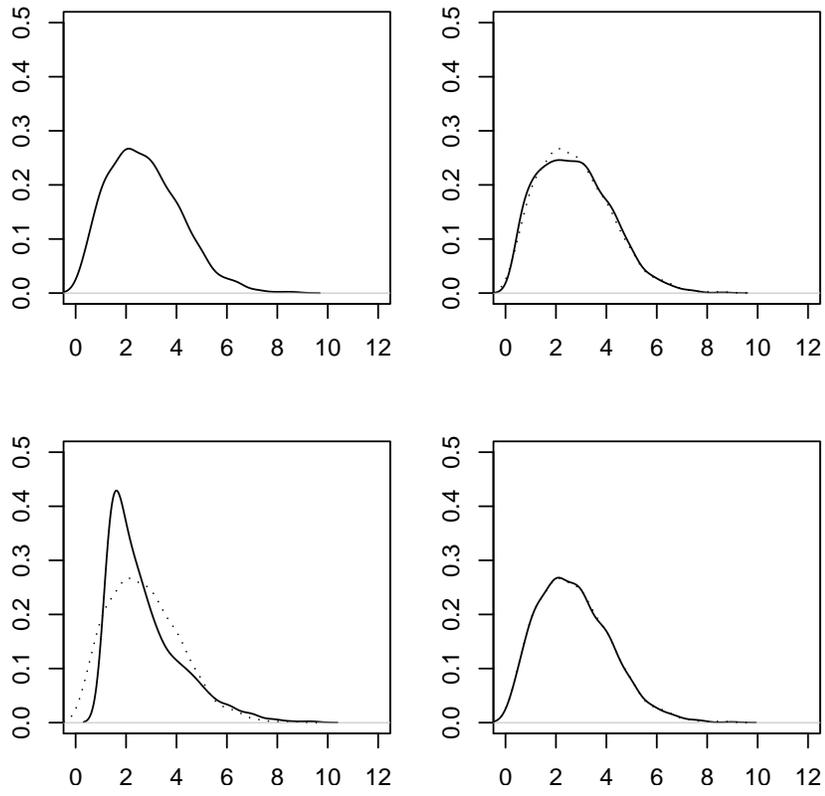}}
  \end{center}
 \caption{
As for figure~\ref{fig8} but for predicted uncertainties calibrated using a shift and a scaling.
The distribution in the top right hand panel is more or less correct now.
}
\label{fig10}
\end{figure}

We then use a more complex calibration system that corrects the predicted uncertainty using a shift
and a scaling. Because such a calibration system has two parameters it can (and does)
correct both the mean and the variance of the uncertainty (as is shown in the top right panel of figure~\ref{fig10})
although it cannot, of course, improve the correlation.
A scatter plot of the results of this calibration step are shown in the upper left panel of figure~\ref{fig7}.

\subsection{Temporal variation of the forecast uncertainty predicted from the second singular vector}

Finally we consider the temporal variations in uncertainty predicted using only the second singular vector.
The correlation between the actual and the predicted uncertainty is close to zero: values from our
3 independent experiments are shown in table~\ref{table2}, and in the upper right panel of figure~\ref{fig5}.
The distribution of predicted uncertainty is shown in figure~\ref{fig8}, lower left panel.

Calibration using a single scaling again does not succeed in correcting both the mean and the variability
of the uncertainty, and again the ratio of the variability of the predicted uncertainty to the real uncertainty 
is too high, as
can be seen from the calibrated distribution in figure~\ref{fig9}.
In fact, the overestimation of the variability of the predicted uncertainty is 
considerably higher than when using the first singular vector.
Scatter plots of the results from calibrating with a simple scaling, and with a shift and scaling, 
are shown in the top right panels of figure~\ref{fig6} and~\ref{fig7}. The effect of a shift and scaling
on the distribution are shown in figure~\ref{fig10}.

\section{Discussion}
\label{discuss}

We have investigated how singular vectors can be used to predict the evolution of errors in a 
simple two dimensional linear stochastic system. The system is designed to mimic the development of errors
in forecasts of the real atmosphere, and the use of singular vectors is designed to mimic the 
use of singular vectors in the ECMWF ensemble prediction system. 

We generate (or rather \emph{define}) the real uncertainty in our system by using initial conditions sampled
from a bivariate normal distribution. We then use this real uncertainty as a basis for comparison for
uncertainty predicted using singular vectors.

Our first singular vector system uses both singular vectors to predict the uncertainty. The predictions
are very successful, and show a 100\% correlation with the actual uncertainty. However, they still need
calibration. A simple calibration consisting of a scaling is sufficient to match both the mean and the
variability of the predicted uncertainty to reality.

Our second singular vector system uses only the leading singular vector to predict the uncertainty.
These predictions are designed to mimic the ECMWF system, which uses a truncated set of singular vectors to
predict forecast uncertainty. The predictions we generate from the truncated system are less successful
and do not have 100\% correlation with the actual uncertainty. 
They also overestimate the variability of the uncertainty by about 20\%, even after a "spread-scaling"
calibration step. 
Because of this if we want to calibrate the forecast to have the correct mean and variability in the level
of uncertainty we need to use two parameters.

Our third singular vector system uses only the second singular vector to predict the uncertainty.
In this case the correlation with the real uncertainty is very poor. However, the variability of the
uncertainty is again overestimated even after spread-scaling calibration. 
This shows that overestimation of the variability of the uncertainty
is not caused by the first singular vector per se, but just by the use of only a single singular vector,
whichever it is.

In the cases where the correlation between the predicted and actual uncertainty is less than 100\% 
it is not necessarily the best thing to do to enforce the mean and the variance of the uncertainty
to be correct. In fact in a real forecast system we cannot implement this method for calibration
anyway because we do not know the variability of the real uncertainty. 
An alternative calibration 
method is to calibrate the uncertainty prediction so as to maximise the log-likelihood, as used
in~\citet{jewsonbz03a}. 
We will consider this possibility, and analyse the effect of such a calibration system on the mean and the variability
of the uncertainty in the context of our simple model in a subsequent article.

To the extent that our system captures some of the dynamics of the full ECMWF forecast system, we conclude
that the use of truncated singular vectors is one reason why~\citet{jewsonbz03a} have found 
that the predictions of uncertainty 
from that system need calibration using third generation calibration models
that treat the mean of the uncertainty and the variability of the uncertainty separately. 
It also suggests that forecasts from the ECMWF
system calibrated using second generation calibration models will tend to overestimate 
the variability of the uncertainty and overpredict extreme events.

One idea that arises from this work is that it might be worth trying to calibrate the
uncertainty of ensemble forecasts using a standard CDF-based distribution transform, which would convert the 
spread of the forecast ensemble to a prediction of the uncertainty, and enforces a sensible
distribution for the latter. The calibration could fix the parameters of the predicted distribution
using maximum likelihood. This might be a better calibration model than the spread regression model
of~\citet{jewsonbz03a}, which only considers predictions of the uncertainty based on linear 
transformations of the ensemble standard deviation. 

\section{Legal statement}

The lead author was employed by RMS at the time that this article was written.

However, neither the research behind this article nor the writing of this
article were in the course of his employment,
(where 'in the course of his employment' is within the meaning of the Copyright, Designs and Patents Act 1988, Section 11),
nor were they in the course of his normal duties, or in the course of
duties falling outside his normal duties but specifically assigned to him
(where 'in the course of his normal duties' and 'in the course of duties
falling outside his normal duties' are within the meanings of the Patents Act 1977, Section 39).
Furthermore the article does not contain any proprietary information or
trade secrets of RMS.
As a result, the lead author is the owner of all the intellectual
property rights (including, but not limited to, copyright, moral rights,
design rights and rights to inventions) associated with and arising from
this article. The lead author reserves all these rights.
No-one may reproduce, store or transmit, in any form or by any
means, any part of this article without the author's prior written permission.
The moral rights of the lead author have been asserted.

\bibliography{jewson}

\newpage
\begin{table}[!h]
\begin{center}
\begin{tabular}{lll}
 expt   & correlation  & sd ratio \\ 
 1 &    0.95 & 1.22 \\ 
 2 &    0.95 & 1.20 \\ 
 3 &    0.96 & 1.20 \\
 \\
\end{tabular}
\end{center}
\caption{
The second column shows the correlation between the uncertainty predicted using the first singular vector 
and the real uncertainty. The third column shows the overestimation of the uncertainty 
after calibration using spread-scaling (where the real uncertainty is 1). 
The three rows show three independent numerical experiments.
}
\label{table1}
\end{table}

\begin{table}[!h]
\begin{center}
\begin{tabular}{lll}
 expt   & correlation  & sd ratio \\ 
 1 &    -0.03 & 1.89 \\ 
 2 &     0.04 & 1.99 \\ 
 3 &    -0.02 & 1.79 \\
 \\
\end{tabular}
\end{center}
\caption{
As for table~\ref{table1} but for the uncertainty predicted using the second singular vector.
}
\label{table2}
\end{table}

\clearpage

\clearpage

\clearpage

\clearpage

\clearpage

\clearpage

\clearpage

\clearpage

\clearpage

\clearpage

\end{document}